# Classifying superconductivity in Moiré graphene superlattices

E.F. Talantsev[1,2]*, R.C. Mataira[3] and W.P. Crump[3,4,5]

[1]M.N. Miheev Institute of Metal Physics, Ural Branch, Russian Academy of Sciences, 18, S. Kovalevskoy St., Ekaterinburg, 620108, Russia

[2]NANOTECH Centre, Ural Federal University, 19 Mira St., Ekaterinburg, 620002, Russia

[3]Robinson Research Institute, Victoria University of Wellington, 69 Gracefield Road, Lower Hutt, 5040, New Zealand

[4]MacDiarmid Institute for Advanced Materials and Nanotechnology, P.O. Box 33436, Lower Hutt 5046, New Zealand

[5]Aalto University, Foundation sr, PO Box 11000, FI-00076 AALTO, Finland

*Corresponding author: E-mail: evgeny.talantsev@imp.uran.ru

## Abstract

Several research groups have reported on the observation of superconductivity in bilayer graphene structures where single atomic layers of graphene are stacked and then twisted at angles θ forming Moiré superlattices. The characterization of the superconducting state in these 2D materials is an ongoing task. Here we investigate the pairing symmetry of bilayer graphene Moiré superlattices twisted at $\theta = 1.05°$, $1.10°$ and $1.16°$ for carrier doping states varied in the range of $n = 0.5 - 1.5 \cdot 10^{12} cm^{-2}$ (where superconductivity can be realized) by analyzing the temperature dependence of the upper critical field $B_{c2}(T)$ and the self-field critical current $J_c$(sf,$T$) within currently available models for single- and two-band $s$-, $d$-, $p$- and $d+id$-wave gap symmetries. Extracted superconducting parameters show that only $s$-wave and a specific kind of $p$-wave symmetries are likely to be dominant in bilayer graphene Moiré superlattices. More experimental data is required to distinguish between the $s$- and remaining $p$-wave symmetries as well as the suspected two-band superconductivity in these 2D superlattices.

## Introduction

For an isotropic, spherical Fermi surface the density of states at the Fermi level is given by:

$$D(E_F) = \frac{8\pi}{h^3} \cdot (m^*)^2 \cdot v_F \tag{1}$$

where $h$ is the Planck constant, $m^*$ is the effective mass of the charge carriers, and $v_F$ is the Fermi velocity. Due to their large effective mass of $m^* \sim 200 \cdot m_e$ (where $m_e$ is the electron mass) heavy fermion superconductors [1] possess a robust superconducting state, characterized by high values of the upper critical field $B_{c2}$ [2] (significantly above the paramagnetic Pauli limit $B_p$) despite a low Fermi velocity, $v_F \sim 5 \cdot 10^3 \, m/s$, in these compounds [2].

On the other hand, Eq. 1 prohibits a superconducting state in materials possessing a charge carrier effective mass of $m^* < 0.1 \cdot m_e$. The is because a large value of $D(E_F)$ requires relativistic values for the Fermi velocity, $v_F \gtrsim 10^8 \, m/s$, which has not yet been observed in any material. In single layer graphene (SLG) and other materials with Dirac cone Fermi surfaces, $D(E_F)$ is given by:

$$D(E_F) = \frac{8 \cdot \pi \cdot |E_F|}{h^2 \cdot v_F^2}, \tag{2}$$

which shows that $D(E_F)$ is inversely proportional to $v_F^2$. Therefore, the prerequisite to convert SLG and other planar honeycomb lattices [3] into intrinsic superconductors is a reduction of $v_F$.

Lopes dos Santos *et al* [4] were the first to understand that $v_F$ can be suppressed in bilayer graphene by rotating the SLG sheets relative to each other by small twist angles θ. Detailed tight-binding model calculations performed by Bistritzer and MacDonald [5] showed that at the Dirac points $v_F$ goes to zero when bilayer graphene is rotated by small angles, creating a Moiré superlattice. These graphene 2D structures are now called magic-angle twisted bilayer graphene (MATBG).

The discovery of intrinsic superconductivity in few-layer stanene (one of the closest analogues of graphene) which possesses a Fermi velocity of $v_F = 4.5 \cdot 10^4 \; m/s$ by Liao *et al* [6] gave clear experimental proof that the suppression of $v_F$ in Dirac cone materials makes it possible to convert these materials into intrinsic superconductors. Several months after Liao's work [6], intrinsic superconductivity in MATBG with $v_F \sim 2 \cdot 10^4 \; m/s$ and $m^* \sim 0.2 \cdot m_e$ was reported by Cao *et al* [7]. There, simultaneous $v_F$ suppression and $m^*$ enhancement was achieved in the way considered by Lopes dos Santos *et al* [4] as well as Bistritzer and MacDonald [5], i.e. by rotation of 2 stacked SLG sheets to form a Moiré superlattice. For instance, Cao *et al* [7] showed that MATBG with $\theta = 1.16°$ exhibits zero resistance at $T_c = 0.14$ K, whereas an angle $\theta = 1.05°$ has zero resistance at $T_c \sim 1.2$ K. More recently, several research groups have discovered a superconducting state in MATBG [8-11], including studies of MATBG at high-pressures (up to $P = 4$ GPa), as well as in trilayer graphene/boron nitride Moiré superlattices [12].

There are still many important questions regarding the superconducting state of MATBG that remain unanswered. In particular, the pairing mechanism and symmetry, as well as the number of superconducting bands, is not clear. While Cao *et al* [7] propose strong electron-electron correlations as the mechanism of the superconductivity, a full analysis with respect to existing theories for phonon mediation is still required. When treated as a phonon mediated superconductor the possibility of multiple superconducting bands arises from the fact that the charge carriers in MATBG experience two different phonon modes, one from intralayer covalent bonds, and a second mode due to the interlayer coupling (details can be found in [13, 14]).

Here, we analyse temperature dependent measurements of the upper critical field and self-field critical current in bilayer twisted SLG structures reported by Cao *et al*. [7] and Lu *et al.* [11]. This analysis is done in the existing phenomenology developed for phonon-mediated superconductors, to test if experimental data necessitates the development of a new phenomenology. We investigate *s*-, *d*-, *p*- and *d+id* pairing symmetry scenarios within single and two-band models. As a result, we found that phonon mediated *d*-wave, *d+id*-wave and three of the four *p*-wave pairing symmetries should be excluded from further consideration. More experimental data is required to confirm which of the remaining two ( i.e., *s*-wave or axial *A⊥l p*-wave) pairing scenarios this material exhibits. As graphene has a planar honeycomb lattice of $sp^2$ bonded carbon atoms, it is unsurprising that phonon-mediated superconductivity would exhibit pairing symmetry of either *s*- or *p*-wave in MATBG.

## Results

### Upper critical field analysis: single superconducting band models

We analyse the temperature dependent perpendicular upper critical field $B_{c2,\perp}(T)$ measured by Cao *et al* [7, Figure 3e] where the magnetic field is applied in the perpendicular direction to the MATBG plane. First we present the standard literature analysis for $B_{c2}(T)$ data by fitting the data to the 5 available models for $B_{c2,\perp}(T)$ and extracting $T_c$ as well as the $\xi(0)$. However, we extend this analysis by noting that the first 5 models presented have an implicit assumption of the behaviour of the penetration depth $\lambda(T)$. By dropping this assumption, we can express $B_{c2,\perp}(T)$ as a function of the band gap $\Delta(T)$ and its symmetry.

To begin, we fit the $B_{c2,\perp}(T)$ data to the Gorter-Casimir (GC) two fluid model [15, 16] which is still in wide use [17,18]:

$$B_{c2,\perp}(T) = \frac{\phi_0}{2 \cdot \pi \cdot \xi_{ab}^2(0)} \cdot \left(1 - \left(\frac{T}{T_c}\right)^2\right), \tag{3}$$

where $\phi_0 = \frac{h}{2 \cdot e} = 2.067 \cdot 10^{-15}$ Wb is the superconducting flux quantum; composed of Plank's constant *h* and the electron charge *e* is, and $\xi_{ab}(T)$ is the coherence length in the a-b plane. The fit is shown in Fig. 1(a).

Another widely used model for fitting $B_{c2}(T)$ data is the Werthamer-Helfand-Hohenberg (WHH) model [19, 20]:

$$B_{c2,\perp}(0) = -0.697 \cdot T_c \cdot \left(\frac{dB_{c2}(T)}{dT}\right)_{T \sim T_c}, \tag{3}$$

a fit of the data to this model is also shown in Fig. 1(a). Baumgartner *et al* [21] proposed a simple and accurate analytical expression that matches the shape of the WHH model:

$$B_{c2,\perp}(T) = \frac{1}{0.697} \cdot \frac{\phi_0}{2 \cdot \pi \cdot \xi_{ab}^2(0)} \cdot \left(\left(1 - \frac{T}{T_c}\right) - 0.153 \cdot \left(1 - \frac{T}{T_c}\right)^2 - 0.152 \cdot \left(1 - \frac{T}{T_c}\right)^4\right). \tag{4}$$

This model will be designated as B-WHH and its fit of the data are shown in Fig. 1(b), where both values of $B_{c2,\perp}(T = 0 \text{ K})$ (i.e., original WHH and B-WHH) agree as expected.

Jones-Hulm-Chandrasekhar (JHC) proposed three models [22], two of which are often used to fit $B_{c2}(T)$ data [23-25]. The first is a combination of Ginzburg-Landau (GL) theory with the Gorter-Casimir expression for the temperature dependence of $\lambda(T)$:

$$B_{c2,\perp}(T) = \frac{\phi_0}{2 \cdot \pi \cdot \xi_{ab}^2(0)} \cdot \left(\frac{1 - \left(\frac{T}{T_c}\right)^2}{1 + \left(\frac{T}{T_c}\right)^2}\right), \tag{5}$$

which we shall refer to as the JHC model, a fit using this model is shown in Fig. 1(c). This model gives a little higher value for $B_{c2,\perp}(T = 0 \text{ K})$ and $T_c$ in comparison with the other models, as well as the lowest value of $\xi_{ab}(0)$. The second model proposed by JHC is based on an expression from Gor'kov for $B_{c2}(T)$ [26]:

$$B_{c2}(T) = B_c(T) \cdot \frac{\sqrt{2}}{1.77} \cdot \frac{\lambda(0)}{\xi(0)} \cdot \left(1.77 - 0.43 \cdot \left(\frac{T}{T_c}\right)^2 + 0.07 \cdot \left(\frac{T}{T_c}\right)^4\right), \tag{6}$$

where $B_c(T)$ is the thermodynamic critical field, and $\lambda(0)$ is the ground state London penetration depth. Jones *et al* [22] again use the Gorter-Casimir form of $\lambda(T)$ to produce a model of the following form:

$$B_{c2,\perp}(T) = \frac{1}{1.77} \cdot \frac{\phi_0}{2 \cdot \pi \cdot \xi_{ab}^2(0)} \cdot \left(1.77 - 0.43 \cdot \left(\frac{T}{T_c}\right)^2 + 0.07 \cdot \left(\frac{T}{T_c}\right)^4\right) \cdot \left(1 - \left(\frac{T}{T_c}\right)^2\right). \tag{7}$$

Eq. 7 will be referred to as the Gor'kov model.

However, rather than assume the behavior of the penetration depth, eq. 6 can be used to build a model based directly on the BCS expression [16,27] for $\lambda(T)$ as a function of the superconducting gap:

$$\lambda(T) = \frac{\lambda(0)}{\sqrt{1 - \frac{1}{2 \cdot k_B \cdot T} \int_0^\infty \frac{d\varepsilon}{\cosh^2\left(\frac{\sqrt{\varepsilon^2 + \Delta^2(T)}}{2 \cdot k_B \cdot T}\right)}}}, \tag{8}$$

where $\Delta(T)$ is the temperature-dependent superconducting gap. Eq. 8 captures the effect of the gap symmetry on $B_{c2}(T)$. An expression for $\Delta(T)$ as a function of the wavefunction symmetry is given by Gross-Alltag $et$ $al$ [28]. For now, we test s-wave symmetry where the gap takes the form:

$$\Delta(T) = \Delta(0) \cdot \tanh\left[\frac{\pi \cdot k_B \cdot T_c}{\Delta(0)} \cdot \sqrt{\eta \cdot \frac{\Delta C}{C} \cdot \left(\frac{T_c}{T} - 1\right)}\right], \tag{9}$$

where $\Delta C/C$ is the relative jump in electronic specific heat at $T_c$, and $\eta = 2/3$. We can then use the standard GL expression:

$$B_{c2}(T) = \sqrt{2} \cdot \frac{\lambda(T)}{\xi(T)} \cdot B_c(T), \tag{10}$$

to restate Gor'kov's expression, eq. 6, as:

$$\kappa(T) = \frac{\lambda(T)}{\xi(T)} = \frac{1}{1.77} \cdot \frac{\lambda(0)}{\xi(0)} \cdot \left(1.77 - 0.43 \cdot \left(\frac{T}{T_c}\right)^2 + 0.07 \cdot \left(\frac{T}{T_c}\right)^4\right). \tag{11}$$

Then, by considering another GL expression:

$$B_{c2}(T) = 2 \cdot \left(\frac{\lambda(T)}{\xi(T)}\right)^2 \cdot \frac{B_{c1}(T)}{\ln(\kappa(T)) + 0.5} = \left(\frac{\lambda(T)}{\xi(T)}\right)^2 \cdot \frac{\phi_0}{2 \cdot \pi \cdot \lambda^2(T)} = \left(\kappa(T)\right)^2 \cdot \frac{\phi_0}{2 \cdot \pi \cdot \lambda^2(T)}, \tag{12}$$

we can combine eqs. (8-11) and eq. 12 into a single expression for the temperature dependent upper critical field:

$$B_{c2,\perp}(T) = \frac{\phi_0}{2 \cdot \pi \cdot \xi_{ab}^2(0)} \cdot \left(\frac{1.77 - 0.43 \cdot \left(\frac{T}{T_c}\right)^2 + 0.07 \cdot \left(\frac{T}{T_c}\right)^4}{1.77}\right)^2 \cdot \left[1 - \frac{1}{2 \cdot k_B \cdot T} \cdot \int_0^\infty \frac{d\varepsilon}{\cosh^2\left(\frac{\sqrt{\varepsilon^2 + \Delta^2(T)}}{2 \cdot k_B \cdot T}\right)}\right]. \tag{13}$$

Eq. 13 is a function parameterized by four fundamental superconducting parameters, $\xi_{ab}(0)$, $\Delta(0)$, $\Delta C/C$, and $T_c$. It is critical to note that the role each parameter plays in the equation is well founded in BCS and GL theory. In particular $\xi_{ab}(0)$ determines the absolute value of $B_{c2,\perp}(0)$, independently of the other parameters. Furthermore, $T_c$ is tightly constrained to define a reduced temperature scale $t = T/T_c$, over which Gor'kov's expression, eq. 6, is valid. The only new feature of eq. 13 is the introduction of the gap symmetry expression for $\lambda(T)$, eq. 8. This introduction, and that of the Gross-Alltag $et.$ $al.$ expression for the gap $\Delta(T)$ itself, only introduces degrees of freedom, $\Delta(0)$ and $\Delta C/C$, whose behavior were phenomenologically assumed by the literature models. The advantage of this method is that the two new parameters can be deduced and explicitly checked against the bounds of the theory's validity.

A fit of the MATBG $B_{c2}(T)$ data to Eq. 13 is shown in Fig. 1(e) where the superconducting gap ratio was found to be:

$$\frac{2 \cdot \Delta(0)}{k_B \cdot T_c} = 3.73 \pm 0.28 \tag{14}$$

and the relative jump in electronic specific heat:

$$\left.\frac{\Delta C}{C}\right|_{T \sim T_c} = 1.46 \pm 0.28. \tag{15}$$

These deduced parameters, within their uncertainties, are remarkably close to the BCS weak-coupling limits of $\frac{2 \cdot \Delta(0)}{k_B \cdot T_c} = 3.53$ and $\left.\frac{\Delta C}{C}\right|_{T \sim T_c} = 1.43$. This is the first quantitative evidence that $intrinsic$ $superconductivity$ in MATBG can be understood in the existing phenomenology of $s$-wave electron-phonon mediated superconductivity and a new phenomenology may not need to be developed. Overall, the deduced values for $\xi_{ab}(0)$ and $T_c$ using the six models are (see for details Supplementary Table S1): $\xi_{ab}(0) = 61.4 \pm 1.7 \; nm$ $T_c = 0.500 \pm 0.006 \; K$.

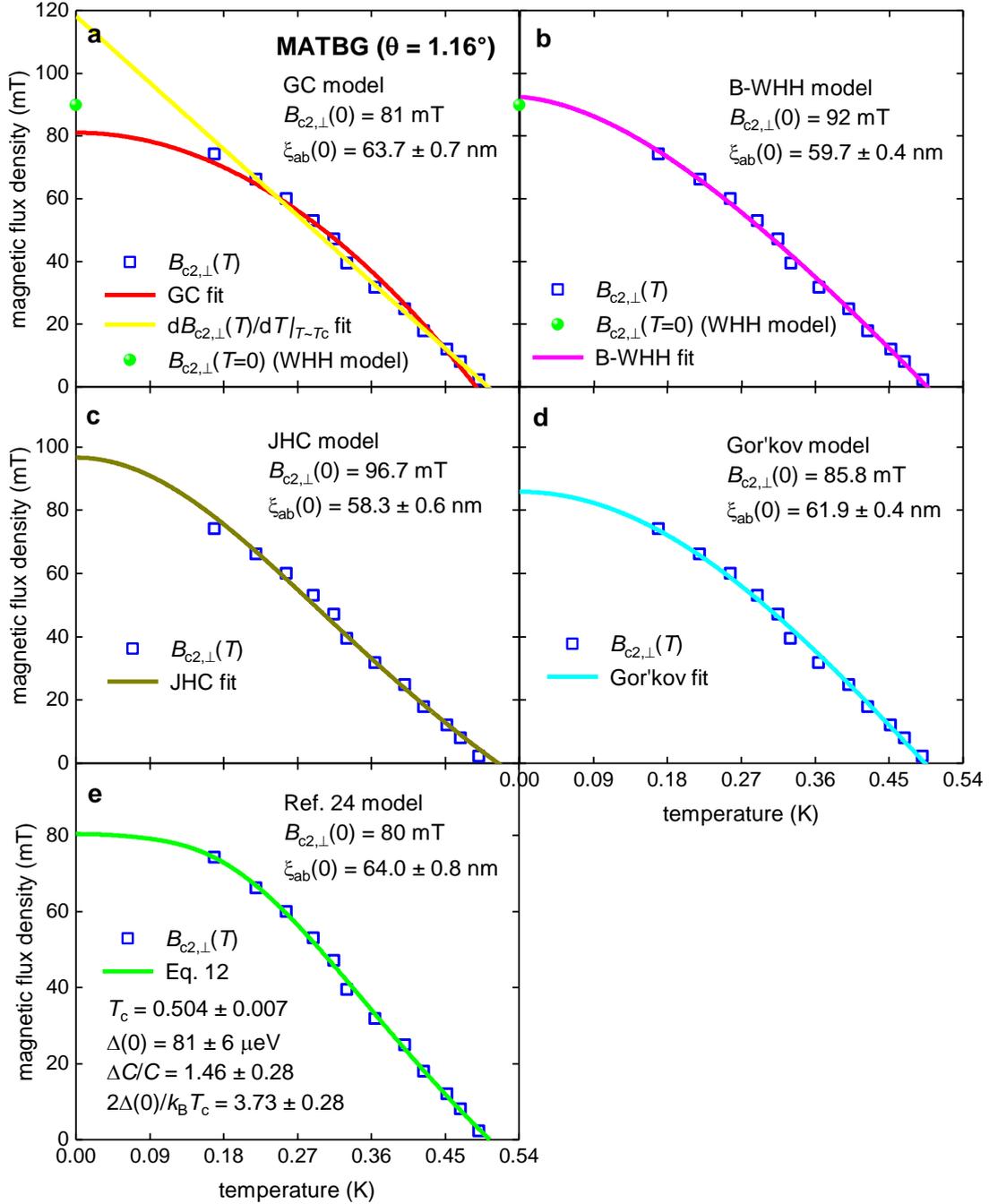

**Figure 1.** The upper critical field data, $B_{c2,\perp}(T)$, for sample M1 ($\theta = 1.16°$) of measured by Cao et al [7] (squares) fitted to single band models. (a) red: fit to Gorter-Casimir (GC) two fluid model (Eq. 1) with a goodness of fit $R = 0.9892$; yellow: Werthamer-Helfand-Hohenberg (WHH) model (Eq. 2), $R = 0.9976$. (b) Baumgartner-WHH (B-WHH) model (Eq. 3), $R = 0.9966$. (c) Jones-Hulm-Chandrasekhar (JHC) model Eq. 4, $R = 0.9930$. (d) Gor'kov model Eq. 6, $R = 0.9959$. (e) Model proposed in Ref. 23, $R = 0.9979$.

Once a value of the ground state gap has been obtained another standard BCS expression [16, 27] can be used to calculate $v_F$ for the sample M1 ($\theta = 1.16°$) based on the deduced values of $\xi_{ab}(0)$ and $\Delta(0)$:

$$v_{F,ab} = \frac{2 \cdot \pi^2 \cdot \Delta(0) \cdot \xi_{ab}(0)}{h} = (2.37 \pm 0.18) \cdot 10^4 \ m/s \tag{16}$$

The value given in Eq. 16 is about two orders of magnitude lower than the Fermi velocity of pure SLG ($v_{F,ab} \sim 10^6 \ m/s$) [29].

The Fermi temperature, $T_F$, can be calculated in the usual way:

$$T_F = \frac{E_F}{k_B} = \frac{m^* \cdot v_F^2}{2 \cdot k_B}, \tag{17}$$

where $m^*$ is the effective mass of charge carriers. In the original paper by Cao et al [7], $m^*$ was not measured for sample M1 ($\theta = 1.16°$). A simple assumption is that at the same doping, MATBG with different $\theta$ should have the same effective mass, $m^*$. By assuming this, and using $B_{c2}(T)$ as an indication of the doping state we can assume that for sample M1:

$$m^*(n \sim -1.5 \cdot 10^{12} cm^{-2}) \approx 0.2 \cdot m_e. \tag{18}$$

(we present full analysis for this value for sample M2 in the following section).

Corresponding to this $m^*$ value the Fermi temperature, $T_F$, is:

$$T_F = 3.7 \pm 0.5 \ K \tag{19}$$

We plot this data point for MATBG sample M1 in an Uemura plot [30] (Fig. 2) where the transition temperature $T_c$ is defined as the temperature at which order parameter phase coherence is established i.e. when the sample resistance becomes zero, $R = 0 \ \Omega$ (rather than R= $0.5R_n$). This definition is in accordance with all other data points (for other materials) and follows the general logic of the Uemura *et al* [30,31] definition of the transition temperature.

From $R(T)$ data for sample M1 presented in [7, Figure. 1b] we determine the transition temperature to be $T_c = 0.136$ K. One therefore obtains:

$$\frac{T_c}{T_F} = 0.037 \pm 0.005 < 0.05, \tag{20}$$

which fits within the boundary of:

$$0.01 \leq \frac{T_c}{T_F} < 0.05, \tag{21}$$

established by Uemura *et al.* [30,31] for all other unconventional superconductors.

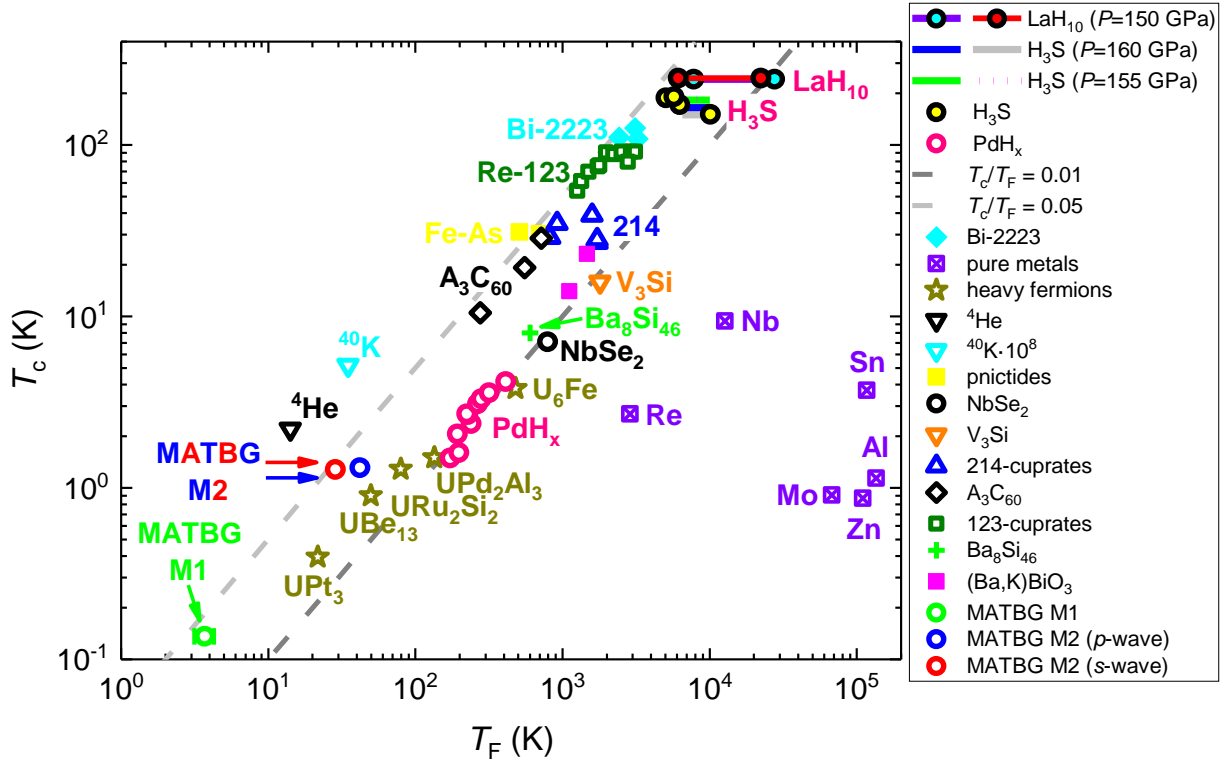

**Figure 2.** A plot of $T_c$ versus $T_F$ obtained for materials representative of the various superconducting families. Data was taken from Uemura [28], Cao *et al* [7], as well as [24,25].

Unfortunately, the existing data set is insufficient to reveal irreconcilable differences between the models analyzed. While this is reasonable considering how low the temperature must be taken, it is also unfortunate that $B_{c2,\perp}(T)$ data is not available for sample M2; where the higher critical temperature would lead to a lower acheivable range of reduced temperature $t = T / T_c$.

It is possible to fit the $B_{c2,\perp}(T)$ data to a model with *d*-, *p*- and *d+id*-gap symmetry by substituting the relevant expressions for $\lambda(T)$ and $\Delta(T)$ in Eq. 13, however, the $B_{c2,\perp}(T)$ data for the sample M1 [7] was not measured to low enough temperatures ($T / T_c < 0.33$) to accurately determine the fit parameters.

To resolve this issue we look at self-field critical current density data $J_c(\text{sf},T)$ which was measured by Cao *et al* [7] for the sample M2 ($\theta = 1.05°$) down to $T = 0.05$ K. These measurements, taking into account that $T_c \sim 1.2$ K, were taken down to a reduced temperature of $T/T_c = 0.04 << 0.33$. Therefore, the raw $J_c(\text{sf},T)$ data provides a meaningful insight into the superconducting gap symmetry of MATBG, which we will explore in a later section.

**Upper critical field analysis: two superconducting band model**. As we already mentioned above, several authors have proposed a two-band superconducting state in MATBG which originates from two different phonon modes. One is due to intralayer covalent bonds, and the other is due to interlayer coupling (an extended reference list and discussion can be found in Refs. 13,14). It is interesting to note that in our previous papers [17,32-34] we found that many atomically thin superconductors, including SLG and topological insulators (with a proximity induced superconducting state) have a two-superconducting band state.

Looking closer at the upper critical field, $B_{c2,\perp}(T)$, fit to the single band GC model in Fig. 3 (a), it can be seen that in the temperature range of $T = 0.3 – 0.4$ K the experimental $B_{c2,\perp}(T)$ data is well below the fitted curve. A fit of the data to a decoupled-two-band GC model (which we proposed in Refs. 17):

$$B_{c2,\perp}(T) = B_{c2,\perp,Band1}(T) + B_{c2,\perp,Band1}(T) \,, \tag{22}$$

is shown in Fig. 3 (b). The deduced parameters, see Fig. 3 (b), agree well with the mutual parameter interdependence criteria, which remains low and varies from 0.69 to 0.97 (the definition of this parameter can be found elsewhere [17]). Giving strong evidence that MATBG is a two band superconductor.

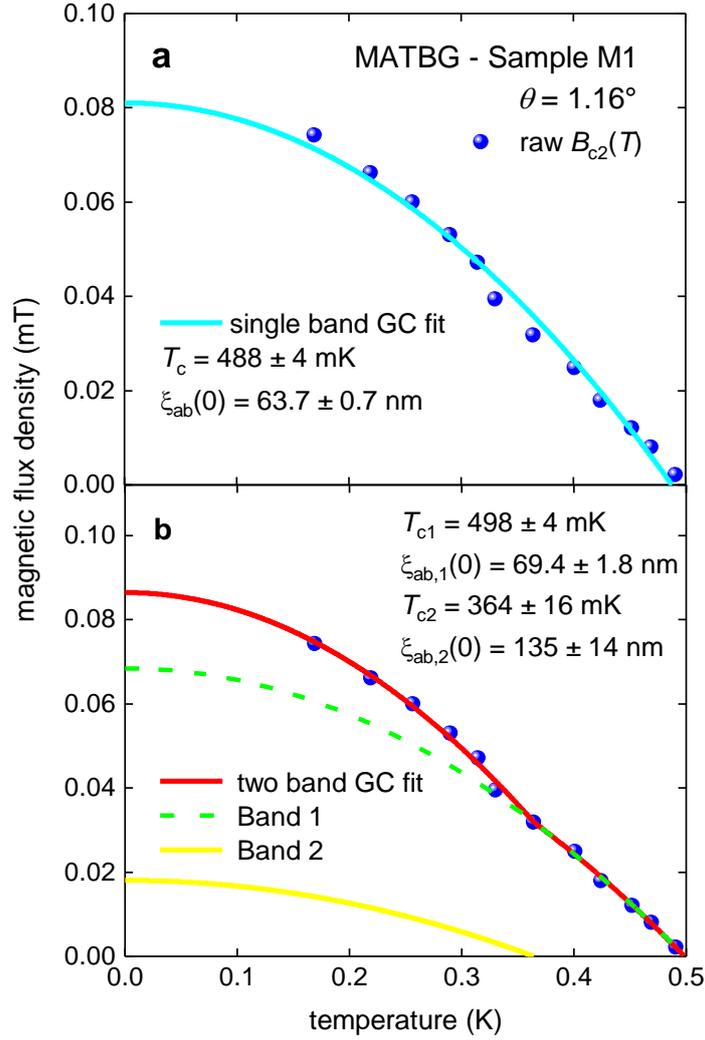

**Figure 3.** The upper critical field data, $B_{c2,\perp}(T)$, for sample M1 ($\theta = 1.16°$) of measured by Cao et al [7] (balls) fitted to two GC two-band models. (a) fit to single band GC model (Eq. 1) with a goodness of fit $R = 0.9892$; (b) fit to two-band GC model (Eq. 22) [17], $R = 0.9984$.

It would be interesting to analyse the experimental $B_{c2,\perp}(T)$ data using the models in the previous section, however this requires a more comprehensive raw $B_{c2,\perp}(T)$ data file which densely covers the reduced temperature range of the lower band (band 2) $t_2 = T/T_{c2}$ down to a level $t << 0.3$. This experimental data is not yet available.

**Self-field critical current analysis: single superconducting band models**. The voltage-current, $V(I)$, curves for sample M2 can be found in [7, Figure 1e]. To extract $I_c(\text{sf},T)$ the experimental $V(I)$ curves are fitted by the standard power-law expression [35-37]:

$$V(I) = V_0 + k \cdot I + V_c \cdot \left(\frac{I}{I_c}\right)^n \tag{23}$$

where $V_0$ is an instrumental offset, $k$ is a linear term used to accommodate incomplete current transfer in short samples, $n$ is the flux creep exponent, and $V_c$ is the voltage criteria which was chosen to be 10 μV. The superconducting state can be defined through the $V(I)$ fit to Eq. 23 by a $n > 1$ criterion, where the normal state corresponds to $n = 1$ (Ohm's law). Physically speaking, this approach defines detection of the superconducting phase coherence by detection of flux vortex flow in the sample. This distinction is necessary as data in Cao *et. al.* shows an elongated transition from the normal to superconducting state [7, Figure 1b]. Furthermore, this definition of $T_c$ is in full accordance with accepted standard in the field [38-43].

We find that $n(T=1.07 \text{ K}) = 2.0 \pm 0.2$, where as $n(T=1.26 \text{ K}) = 1.1 \pm 0.7$. The bridge resistance is found to be $R(T=1.26 \text{ K}) = 1.56 \text{ k}\Omega$, with a material resistivity $\rho(T=1.26$ K)$=7.5\cdot10^{-7}$ Ω·m (the MATBG thickness, $2b$, was assumed to be 1.0 nm, i.e. around double the lower limit of the most accurate measurements of single layer graphene [44], $d$=0.43-1.69 nm). By eq. 23, the highest temperature at which the superconducting coherence was confirmed in the experiment is $T = 1.07$ K (see, Figs. S1, S2, S3, and Table S2).

The resulting $J_c(\text{sf},T) = I_c(\text{sf},T)/(4ab)$ (where $2a$ is the sample M2 width, $2a = 1.05$ μm) is shown in Fig. 4. In this figure the $J_c(\text{sf},T)$ data was fitted using different superconducting gap symmetries by the equation [45]:

$$J_c(sf,T) = \frac{\phi_0}{4\cdot\pi\cdot\mu_0} \cdot \frac{ln\left(\frac{\lambda_{ab}(T=0)}{\xi_{ab}(T=0)}\right)+0.5}{\lambda_{ab}^3(T)} \tag{24}$$

where $\mu_0$ is the magnetic permeability of free space. Eq. 24 is valid for thin superconductors if $2b < \lambda(0)$ [17,18,45-47] and this should be the case for 1.0 nm thick MATBG.

Unfortunately, raw $B_{c2,\perp}(T)$ experimental data, from which to deduce $\xi_{ab}(0)$ for sample M2, is unavailable. However, it can be seen that $\xi_{ab}(0)$ for both sample M1 and sample M2 is practically identical if we consider values at optimal doping. This follows by inspection of $B_{c2,\perp}(T)$ in [7, Figure 3e] for sample M1, where the comparison of $B_{c2,\perp}(T)$ for both samples is displayed in [7, Figure 3a] and [7, Figure 3b]. Thus, for the analysis of sample M2 by Eq. 23 we will use the value of $\xi_{ab}(0) = 61.4$ nm deduced from sample M1. This assumption is also supported by a recent report from Lu *et al.* [11] (in their Extended Data Figure 3), who measured $B_{c2}(T=16$ mK$)$ for a broad range of doping states for MATBG with $\theta \sim 1°$. We analyse the Lu *et al.* [11] data in the following section.

There are several recent first principle calculation papers where different types of superconducting gap symmetries are proposed to present in MATBG. For instance, *d-*, *p-*, and exotic *d+id*-wave symmetries, where *d+id*-wave symmetry was initially considered by Laughlin [48] twenty years ago for HTS cuprates. We only mention [13,14,49-56] where models and extended reference lists can be found.

However, of more importance is whether such gap symmetries are supported by experimental data. Therefore, we fit the available data using an extended BCS model with different expressions for the gap symmetry and compare the deduced parameters with weak-coupling BCS limits. Based on this we can infer which gap symmetries can potentially explain the measured behaviour. Here, we fit the experimental $J_c(\text{sf},T)$ data for sample M2 to Eq. 24 by utilizing different expressions for the symmetry of the superconducting gap. For these expressions, we again draw from the approach proposed by Gross-Alltag *et al* [28] for *s-*, *d-*, and *p*-wave symmetries. For *d+id* symmetry we use an approach proposed by Pang *et al* [57].

Equations for $\lambda(T)$ and $\Delta(T)$ for *s*-wave symmetry have already been presented in Eqs. 8,9 respectively, and in Fig. 4(a) a fit of the $J_c(\text{sf},T)$ data to single band *s*-wave model is shown. All the deduced parameters are presented in Table I.

Next, we performed a fit to a *d*-wave gap symmetry model. Using a 2D cylindrical Fermi surface the equation for the London penetration depth is as follows:

$$\lambda(T) = \frac{\lambda(0)}{\sqrt{1 - \frac{1}{2 \cdot k_B \cdot T} \cdot \int_0^{2\pi} cos^2(\theta) \cdot \left( \int_0^{\infty} \frac{d\varepsilon}{cosh^2\left(\frac{\sqrt{\varepsilon^2 + \Delta^2(T,\theta)}}{2 \cdot k_B \cdot T}\right)} \right) \cdot d\theta}} \tag{25}$$

where the superconducting energy gap, $\Delta(T,\theta)$, is given by:

$$\Delta(T, \theta) = \Delta_m(T) \cdot cos(2\theta) \tag{26}$$

where $\Delta_m(T)$ is the is the maximum amplitude of the $k$-dependent $d$-wave gap given by Eq. 9, $\theta$ is the angle around the Fermi surface subtended at $(\pi, \pi)$ in the Brillouin zone (details can be found elsewhere [58,59]). In Eq. 9 the value of $\eta = 7/5$ [58]. The fit to this model does not converge, as the value $\Delta_m(0)$ tends toward an infinitely large value.

**Table 1.** Deduced parameters for *single-band* models applied to MATBG sample M2 [7] doped at $n_n$=-1.44 $10^{12}$ cm$^{-2}$ where an effective mass of charge carriers $m^*/m_e$=0.1637±0.0154 was used, which was obtained from the analysis presented in Fig. 5. For the case of $d+id$ we fixed the ratio of $\Delta C/C = 0.995$ for both gaps. The ground-state coherence length was assumed to be $\xi(0) = 61.4 \pm 1.7$ nm.

| Model | $T_c$ (K) | | $\Delta(0)$ (μeV) | $2\Delta(0)/k_B T_c$ | $\Delta C/C$ | $v_F$ ($10^4$ m/s) | $T_F$ (K) | $T_c/T_F$ | $\lambda(0)$ (nm) |
|---|---|---|---|---|---|---|---|---|---|
| *s*-wave | 1.28 ± 0.05 | | 245 ± 9 | 4.4 ± 0.2 | 2.7 ±0.9 | 7.3 ± 0.3 | 28.6 ± 2.0 | 0.045 ± 0.002 | 2,182 ± 3 |
| *d*-wave (low-*T* asymptote) | | | 2,300 ± 500 | > 40 | | | | | 2,166 ± 5 |
| *p*-wave polar **A**⊥*l* | 1.24 ± 0.06 | | >$10^6$ | > $10^4$ | >7 | | | | 2,142 ± 25 |
| *p*-wave polar **A**∥*l* | 1.34 ± 0.05 | | 430 ± 32 | 7.4 ± 0.6 | 1.0 ±0.3 | 13 ± 1 | 86 ± 15 | 0.015 ± 0.003 | 2,180 ± 3 |
| *p*-wave axial **A**⊥*l* | 1.31 ± 0.05 | | 301 ± 13 | 5.4 ± 0.2 | 1.9 ±0.6 | 8.8 ± 0.3 | 42 ± 3 | 0.031 ± 0.002 | 2,183 ± 3 |
| *p*-wave axial **A**∥*l* | 1.37 ± 0.05 | | 1,500 ± 1,200 | 27 ± 22 | 2.3 ±0.5 | 46 | 1,100 | >0.001 | 2,178 ± 3 |
| *d+id*-wave | 1.27 ± 0.05 | Gap 1 | 322 ± 38 | 5.92 ± 0.70 | 0.959 (fixed) | 9.5 ± 1.0 | 49 ± 10 | 0.026 ± 0.003 | 2,180 ± 3 |
| | | Gap 2 | 187 ± 24 | 3.44 ± 0.30 | 0.959 (fixed) | 5.5 ± 0.5 | 16 ± 3 | 0.078 ± 0.016 | |

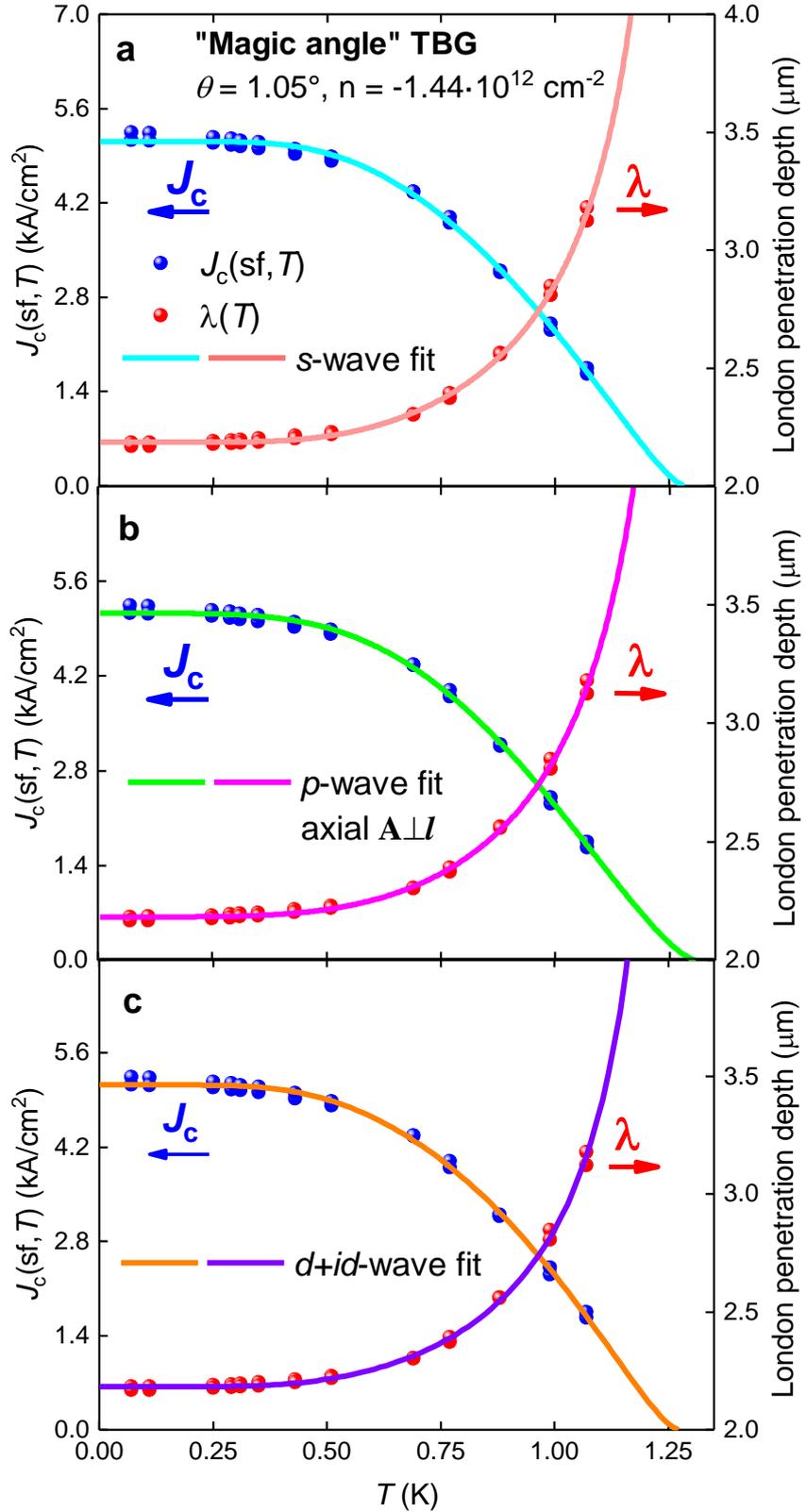

**Figure 4.** The self-field critical current density, $J_c$(sf,$T$), for sample M2 ($\theta = 1.05°$) from the work of Cao et al [7] and a fit of the data to three single-band models. For all models $\xi_{ab}(0) = 61.4$ nm was used. (a) $s$-wave fit, the goodness of fit $R = 0.9964$; (b) $p$-wave axial $\mathbf{A} \perp l$ fit, $R = 0.9971$; (c) $d+id$-wave fit, $R = 0.9967$. Deduced parameters are listed in Table I.

A fit can still be obtained by using a low-$T$ asymptote of the single-band $d$-wave model which is presented in Fig. S4 It can be seen (Fig. S4, Table I) that the deduced $\Delta_m(0) = 2.3 \pm 0.5$ meV is unacceptably large. These results indicate that phonon-mediated $d$-wave symmetry in MATBG is not supported by experimental data and should be omitted from further consideration.

Fitting a $p$-wave gap symmetry model is more complicated (compared with $s$- and $d$-wave) because in this case the gap function is given by [28]:

$$\Delta(\widehat{\boldsymbol{k}}, T) = \Delta(T) f(\widehat{\boldsymbol{k}}, \widehat{\boldsymbol{l}}) \tag{27}$$

where, $\Delta(T)$ is the superconducting gap amplitude, $\boldsymbol{k}$ is the wave vector, and $\boldsymbol{l}$ is the gap axis. The electromagnetic response depends on the mutual orientation of the vector potential and the gap axis. In an experiment this is given by the orientation of the crystallographic axes compared with the direction of the electric current. There are two different $p$-wave pairing states: "axial" where there are two point nodes, and "polar" where there is an equatorial line node. The shape of the London penetration depth, $\lambda(T)$ for $p$-wave polar $\mathbf{A}\|\boldsymbol{l}$ and axial $\mathbf{A}\perp\boldsymbol{l}$ cases are difficult to distinguish from the $s$-wave counterpart, and the $p$-wave axial $\mathbf{A}\|\boldsymbol{l}$ case is difficult to distinguish from the dirty $d$-wave case [28,59].

Despite these difficulties there is still a possibility to make a distinction based on the values of the deduced fundamental superconducting parameters, in particular by considering the ratios of $\frac{2\cdot\Delta(0)}{k_B\cdot T_c}$ and $\Delta C/C$. These two ratios are given in Table 2 for $p$-wave and other gap symmetries [28].

**Table 2.** BCS weak-coupling limit values for $2\Delta(0)/k_B T_c$ and $\Delta C/C$ for $s$-, $d$-, $p$-, and $d+id$-wave superconducting gap symmetries [28,59]. $\Delta_m(0)$ is the maximum amplitude of the $k$-dependent $d$-wave gap, $\Delta(\theta) = \Delta_m(0) \cdot \cos(2\theta)$.

| Pairing symmetry and experiment geometry | $\frac{2 \cdot \Delta(0)}{k_B \cdot T_c}$ or $\frac{2 \cdot \Delta_m(0)}{k_B \cdot T_c}$ | $\frac{\Delta C}{C}$ |
|---|---|---|
| $s$-wave | 3.53 | 1.43 |
| $d$-wave | 4.28 | 0.995 |
| $p$-wave; polar $\mathbf{A} \perp \mathbf{l}$ | 4.924 | 0.792 |
| $p$-wave; polar $\mathbf{A} \parallel \mathbf{l}$ | 4.924 | 0.792 |
| $p$-wave; axial $\mathbf{A} \perp \mathbf{l}$ | 4.058 | 1.188 |
| $p$-wave; axial $\mathbf{A} \parallel \mathbf{l}$ | 4.058 | 1.188 |
| $d+id$ | 3.85 | depends on the ratio of two gap amplitudes; $0.995 \leq \frac{\Delta C}{C} \leq 1.43$ |

A gap equation for the $p$-wave case was given by Gross-Alltag *et. al.* [28] which is identical to Eq. 9, but $\eta$ is given by:

$$\eta_{p,a} = \frac{2}{3} \cdot \frac{1}{\int_0^1 f_{p,a}^2(x) \cdot dx} \tag{28}$$

where

$$f_p(x) = x \; ; \text{polar configuration} \tag{29}$$

$$f_a(x) = \sqrt{1-x^2} \; ; \text{axial configuration} \tag{30}$$

This gap equation was substituted into the temperature dependent London penetration depth equation given also by Gross-Alltag *et al* [28]:

$$\lambda_{(p,a)(\perp,\parallel)}(T) = \frac{\lambda_{(p,a)(\perp,\parallel)}(0)}{\sqrt{1 - \frac{3}{4 \cdot k_B \cdot T} \cdot \int_0^1 w_{\perp,\parallel}(x) \cdot \left( \int_0^\infty \frac{d\varepsilon}{\cosh^2\left( \frac{\sqrt{\varepsilon^2 + \Delta_{p,a}^2(T) \cdot f_{p,a}^2(x)}}{2 \cdot k_B \cdot T} \right)} \right) \cdot dx}} \tag{31}$$

where the function $w_{\perp,\parallel}(x)$ is $w_\perp(x) = (1 - x^2)/2$ and $w_\parallel(x) = x^2$.

By substituting Eqs. 9, 28-31 in Eq. 24, one can fit the experimental $J_c$(sf,$T$) data to the polar and axial $p$-wave model and deduce $\lambda(0)$, $\Delta(0)$, $\Delta C/C$ and $T_c$ as free-fitting parameters.

The result for only one *p*-wave configuration (axial $A \perp l$) is shown in Fig. 4(b) as this has the only meaningfully deduced parameters. The other three fits are presented in Figs. S5. The deduced parameters for all cases are presented in Table 1.

The gap equation for the case of *d+id* gap symmetry is considered elsewhere [48,49,53,56]. In this paper we use the approach proposed by Pang *et al* [57]:

$$\Delta(T,\theta) = \left[\left(\Delta_1(0) \cdot cos(2 \cdot \theta)\right)^2 + \left(\Delta_2(0) \cdot sin(2 \cdot \theta)\right)^2\right]^{1/2} \cdot \left(tanh\left[\frac{2 \cdot \pi}{\alpha_{\text{BCS},d+id-\text{wave}}} \cdot \right.\right.$$

$$\left.\left.\sqrt{\eta \cdot \frac{\Delta C}{C} \cdot \left(\frac{T_c}{T} - 1\right)}\right]\right), \tag{32}$$

where $\Delta_1(0)$ and $\Delta_2(0)$ are the two *d*-wave gap amplitudes, $\eta = 7/5$, and $\theta$ is the angle around the Fermi surface subtended at $(\pi, \pi)$ in the Brillouin zone, and $\alpha_{\text{BCS,d+id-wave}}$ is the double band gap ratio (details can be found elsewhere [56,57]).

Because the experimental $J_c(\text{sf},T)$ dataset was not dense and the two gap model has many parameters, we were forced to reduce the number of free-fitting parameters so as not to overfit the data. We therefore have chosen to fix:

1. $\Delta C/C$ was set to the weak-coupling limit of BCS theory for *d*-wave superconductors, $\Delta C/C = 0.959$.

2. The double band gap ratio $\alpha_{\text{BCS},d+id-\text{wave}}$ was set to its weak-coupling limit which is 1.925.

The fit values obtained for this *d+id* model (Table 1) give the two values of the gap ratios:

$$\frac{2 \cdot \Delta_1(0)}{k_B \cdot T_c} = 5.92 \pm 0.70, \tag{33}$$

$$\frac{2 \cdot \Delta_2(0)}{k_B \cdot T_c} = 3.44 \pm 0.30. \tag{34}$$

where $2 \cdot \Delta_1(0)/k_B T_c$ far exceeds the weak coupling limit of 3.85 for *d+id* symmetry superconductor [48]. This therefore suggests that *d+id* symmetry should also be excluded from further consideration.

The analysis of the self-field critical current density in MATBG within a single band model shows that:

1. MATBG has an equal chance to be moderately strong coupled *s*-wave or *p*-wave superconductor.

2. The ground-state London penetration depth is independent of gap symmetry:

$$\lambda(0) = 2{,}183 \pm 3\ nm. \tag{35}$$

3. The GL parameter $\kappa(0)$ is:

$$\kappa(0) = \frac{\lambda(0)}{\xi(0)} = 35.6. \tag{36}$$

The Fermi velocity $v_{ab,F}$ and the Fermi temperature $T_F$ for the sample M2 ($\theta = 1.05°$) based on the deduced values of $\xi_{ab}(0)$ and $\Delta(0)$ for all gap symmetries have been calculated using Eqs. 16 and 17 respectively, values are presented in Table 1.

This is done for MATBG at the doping state of $n$ = -1.440·10$^{12}$ cm$^{-2}$. To do this, the experimental $m^*(n)/m_e$ data presented in of [7, Figure 5e] is linearly extrapolated:

$$m^*(n = -1.440 \cdot 10^{12} cm^{-2}) = (0.1637 \pm 0.0154) \cdot m_e, \tag{37}$$

which is shown in Fig. 5. This value is close to the last experimental data point in [7, Figure 5e]:

$$m^*(n = -1.468 \cdot 10^{12} cm^{-2}) = 0.1653 \cdot m_e. \tag{38}$$

All the calculated $T_F$ values are in presented in Table III together with the ratio $T_c/T_F$. This ratio varied in the range:

$$0.031 \leq \frac{T_c}{T_F} \leq 0.045, \tag{39}$$

with the lower limit corresponding to the *p*-wave fit and upper limit corresponding to the *s*-wave fit. It should be noted that both lower and upper limits of this ratio are within the range (Eq. 21) established for all unconventional superconductors by Uemura et al [30,31].

From the deduced $\lambda(0)$ values (Eq. 35) the Cooper pair density, $n_{s,C}$, in MATBG ($\theta =$ 1.05°) can be calculated giving:

$$n_{s,C} = \frac{1}{2} \cdot \frac{m^*}{\mu_0 \cdot e^2 \cdot \lambda^2(0)} = (4.9 \pm 0.4) \cdot 10^{23} \; m^{-3}. \tag{40}$$

By using a thickness of $2b = 1.0$ nm, the surface pair density is:

$$n_{s,C,surf} = \frac{1}{2} \cdot \frac{m^*}{\mu_0 \cdot e^2 \cdot \lambda^2(0)} \cdot 2b = (4.9 \pm 0.4) \cdot 10^{14} \; m^{-2} = (4.9 \pm 0.4) \cdot 10^{10} \; cm^{-2} \tag{41}$$

From Eqs. 40,41 the ratio of the Cooper pair density to the total carrier density $n_n = -1.44 \cdot 10^{12} cm^{-2}$ is calculated to be

$$\frac{n_{s,C,surf}}{n_n} = 0.034 \pm 0.003. \tag{42}$$

This value is also in the same range as other unconventional superconductors [30,31].

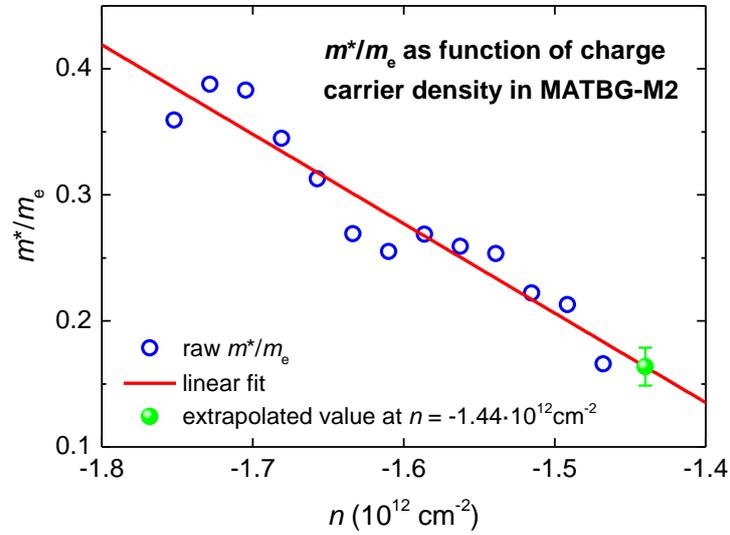

**Figure 5.** Raw experimental $m^*(n)/m_e$ data with a linear fit where the goodness of fit was $R$=0.911. The green spot indicates the extrapolated $m^*(n)/m_e$ at $n$ = -1.44·10$^{12}$ cm$^{-2}$.

**Self-field critical current analysis: two-superconducting band models**. Now we can consider the question: does the available experimental $J_c$(sf,$T$) data support the existence of two-band superconductivity in MATBG, where one band originates from intralayer coupling, and the second from a weak van der Waals interlayer interaction. We fit the $J_c$(sf,$T$) data to a two-band model proposed earlier [17]:

$$J_c(sf, T) = J_{c,band1}(sf, T) + J_{c,band2}(sf, T) \qquad (43)$$

where each band is described by Eq. 24. Because each band has four free-fitting parameters, i.e. $\lambda(0)$, $\Delta(0)$, $\Delta C/C$ and $T_c$, we were forced to reduce the number of parameters, and as we found in [17], the most appropriate approach is to equalize $\Delta C/C$ for both bands, i.e.:

$$\frac{\Delta C_1}{C_1} = \frac{\Delta C_2}{C_2} \qquad (44)$$

We also assume that for both bands $\kappa_c = 35.6$ (Eq. 36). The result of fits to two-band $s$-wave and two-band $p$-wave axial $\mathbf{A} \perp \boldsymbol{l}$ models are shown in Figs. 6(a) and 6(b), respectively. There is experimental evidence that at $T \sim 0.5$ K a new superconducting band opens. The deduced parameters for both fits are in Table 3.

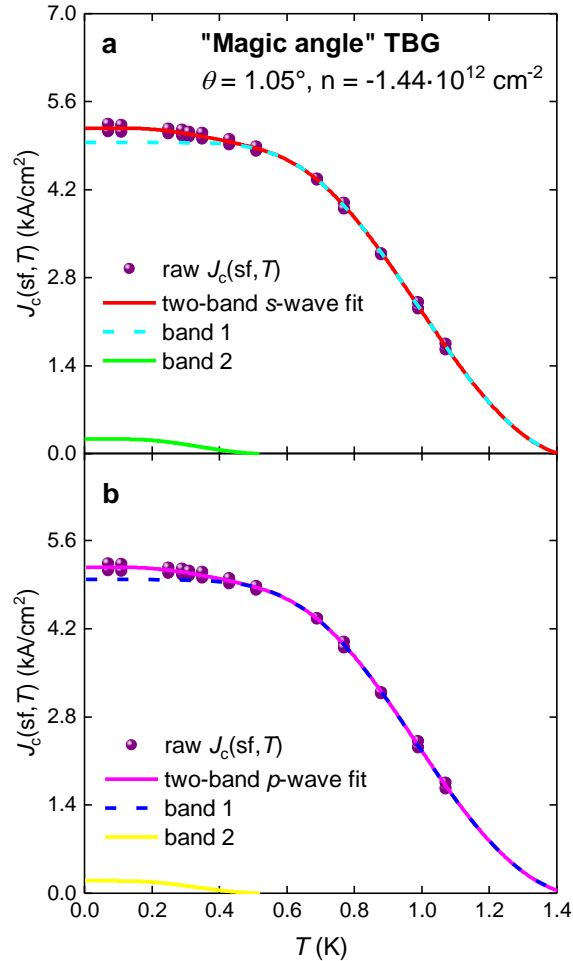

**Figure 6.** The self-field critical current density, $J_c(sf,T)$, for sample M2 ($\theta = 1.05°$) from the work of Cao et al [7] and a fit of the data to $s$- and $p$-wave two-band models. For both models and both bands $\kappa_c(0) = 35.6$ (Eq. 35) was used. (a) $s$-wave fit, the goodness of fit $R = 0.9965$; (b) $p$-wave axial $\mathbf{A} \perp \boldsymbol{l}$ fit, $R = 0.9965$. Deduced parameters are in Table IV.

It can be seen from Fig. 6 and Table 3, that there is not a significant differences between transition temperatures, $T_{c1}$ and $T_{c2}$, of each model. nor is there a significant difference comparing the difference between the BCS ratios of $\frac{2\cdot\Delta_1(0)}{k_B\cdot T_{c1}}$ and $\frac{2\cdot\Delta_2(0)}{k_B\cdot T_{c2}}$ for both bands in both models.

The main difference is in the superfluid densities of the bands, i.e., $\rho_{s,1} \equiv \frac{1}{\lambda_1^2(0)}$ and $\rho_{s,2} \equiv \frac{1}{\lambda_2^2(0)}$, which are different by an order of magnitude. This is a reasonable result given the interlayer charge carrier concentration is much lower in comparison with interlayer one, as the two SGL are only interacting via weak van der Waals forces.

**Table 3.** Deduced parameters for *two-band* models for MATBG sample M2 (Ref. 7) doped at $n_n$=-1.44 $10^{12}$ cm$^{-2}$ where for both bands we assumed an effective mass of charge carriers, $m^*/m_e$=0.1637±0.0154, and $\kappa_c$ = 35.6 (Eq. 36).

| Model | | $T_c$ (K) | $\Delta(0)$ (µeV) | $2\Delta(0)/k_B T_c$ | $\Delta C/C$ | $v_F$ ($10^4$ m/s) | $T_F$ (K) | $T_F/T_c$ | $\lambda(0)$ (nm) | $\dfrac{\rho_{s,band2}}{\rho_{s,band1}}$ |
|---|---|---|---|---|---|---|---|---|---|---|
| Two-band *s*-wave | Gap 1 | 1.42 ± 0.10 | 300 ± 54 | 4.9 ± 0.8 | 1.3 ± 0.7 | 8.8 ± 1.5 | 42 ± 14 | 0.034 ± 0.008 | 2,209 ± 17 | 0.13 ± 0.04 |
| | Gap 2 | 0.5 ± 0.2 | 84 ± 28 | 4.0 ± 1.5 | | 6.8 ± 2.3 | 25 ± 14 | 0.02 ± 0.01 | 6,125 ± 1057 | |
| Two-band axial **A**⊥*l* *p*-wave | Gap 1 | 1.44 ± 0.12 | 390 ± 109 | 6.3 ± 2.0 | 0.9 ± 0.5 | 11.4 ± 2.8 | 70 ± 35 | 0.02 ± 0.01 | 2,204 ± 18 | 0.11 ± 0.04 |
| | Gap 2 | 0.5 ± 0.2 | 98 ± 48 | 4.5 ± 2.5 | | 7.2 ± 3.5 | 28 ± 19 | 0.02 ± 0.01 | 6,467 ± 1385 | |

**MATBG phase diagram.**

Here we use Eq. 23 to analyse data reported by Lu *et al.* [11] on the phase diagram of MATBG with θ ~ 1° measured for a wide range of doping states. We note that Lu *et al.* [11] clearly observe at least six superconducting domes shown in [11, Figure 1,d]. We analyse $B_{c2}$,($T$ = 16 mK) and $I_c$(sf, $T$ = 16 mK) data for the sample D1 which is displayed in Extended Data [11, Figure 3a] and Extended Data [11, Figure 6j-m] respectively.

By using a sample D1 width, $2a = 2$ μm, and thickness, $2b = 1$ nm, we calculate $J_c$(sf, $T=16$ mK) for four doping states where $I_c$(sf, $T=16$ mK) data was reported. By using the deduced $\xi_{ab}(T=16$ mK) for the same doping state we numerically solved Eq. 23 and deduced $\lambda_{ab}(T=16$mK), $\kappa_c(T=16$mK), $n_{s,C,surf}$, and $\frac{n_{s,C,surf}}{n_n}$ (Fig. 7).

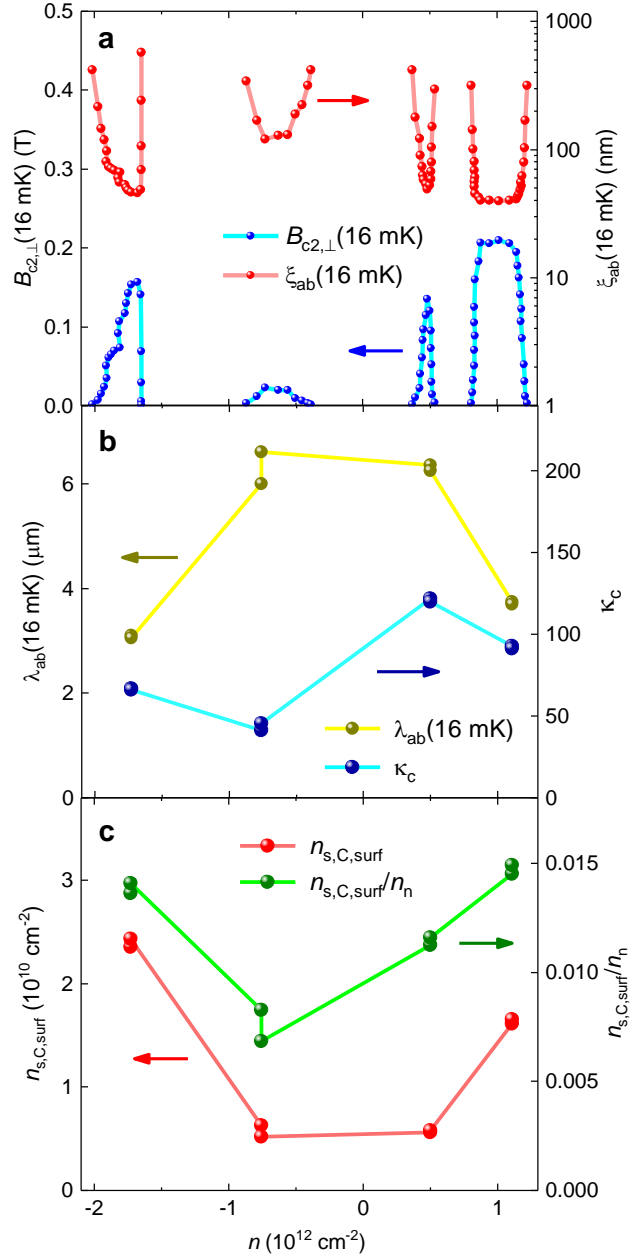

**Figure 7.** Analysis of the superconducting phase diagram of MATBG with θ ~ 1.1°. (a) the upper critical field, $B_{c2,\perp}(16$ mK), and deduced $\xi_{ab}(16$ mK) by Eq. 1. (b) deduced $\lambda_{ab}(16$ mK) and $\kappa_c$ for four doping states for which $I_c$(sf, 16mK) was reported by Lu *et al.* [11]. (c) Cooper pairs surface density, $n_{s,C,surf}$, and the ratio of $\frac{n_{s,C,surf}}{n_n}$ for four doping states for which $I_c$(sf, 16mK) was reported by Lu *et al.* [11].

An interesting result is that for all four different superconducting domes, the ratio of:

$$0.007 \leq \frac{n_{s,C,surf}}{n_n} \leq 0.015 \qquad (45)$$

does not change significantly over the whole phase diagram. What is more surprizing is that two superconducting domes located near $n_n$ = -1.73 $10^{12}$ cm$^{-2}$ and $n_n$ = 1.11 $10^{12}$ cm$^{-2}$ have practically the same $\frac{n_{s,C,surf}}{n_n} \cong 0.015$, and more or less close values for $\lambda_{ab}$($T$=16mK) and $\kappa_c$($T$=16mK).

## IV. Conclusions

In this paper we use existing BCS and GL models to analyze raw experimental data from MATBG reported by Cao *et al.* [7] and Lu *et al.* [11]. Surprisingly enough, the results of our analysis show that MATBG has a very low charge carrier effective mass, $m^* = (0.164 \pm 0.015) \cdot m_e$, and is located in the Uemura plot (Fig. 2), next to the heavy fermion superconductors [1], particularly to UBe$_{13}$ which has $m^* \sim 200 \cdot m_e$ [2]. This places MATBG in the same band of $T_F/T_c$ values as all other unconventional superconductors as categorized by Uemura, contrary to [7, Figure 6].

Our analysis of the temperature dependent upper critical field and the self-field critical current experimental data show that three of four *p*-wave as well as *d-,* and *d+id*-wave symmetries should be excluded from further consideration as possible phonon-electron mediated pairing symmetries in MATBG. Furthermore, the analysis indicates that MATBG is a *moderately strong coupled two-band superconductor* with *s*- or *p*-wave symmetry which can be categorized using the established phenomenology of superconductivity. Because graphene has planar honeycomb lattice of *sp*$^2$ bonded carbon atoms, our findings of either *s*- or *p*-wave pairing symmetry have supporting background evidence.

**Acknowledgements**

Authors thank Mr. Yuan Cao and Prof. P. Jarillo-Herrero (Massachusetts Institute of Technology) for providing raw experimental data and comments on the manuscript and Prof. J.L. Tallon (Victoria University of Wellington) for valued impact and help during manuscript preparation.

EFT thanks financial support provided by the state assignment of Minobrnauki of Russia (theme "Pressure" No. AAAA-A18-118020190104-3) and by Act 211 Government of the Russian Federation, contract No. 02.A03.21.0006.

**Competing interests**

The authors declare no competing interests.

*Authors Contribution*

EFT conceived the work, EFT and RCM performed data analysis, EFT developed critical current data fitting code for $s$-wave pairing symmetry and codes for the upper critical field data fit, WPC developed critical current data fitting codes for $d$-, $p$-, and $d+id$ pairing symmetries, EFT and WPC fitted critical current data, EFT drafted manuscript which was revised by WPC and RCM.

**SUPPLEMENTARY INFORMATION**

**Classifying superconductivity in Moiré graphene superlattices**

E.F. Talantsev[1,2]*, R.C. Mataira[3] and W.P. Crump[4,5]

[1]M.N. Mikheev Institute of Metal Physics, Ural Branch, Russian Academy of Sciences, 18, S. Kovalevskoy St., Ekaterinburg, 620108, Russia

[2]NANOTECH Centre, Ural Federal University, 19 Mira St., Ekaterinburg, 620002, Russia

[3]Robinson Research Institute, University of Wellington, 69 Gracefield Road, Lower Hutt, 5040, New Zealand

[4]MacDiarmid Institute for Advanced Materials and Nanotechnology, P.O. Box 33436, Lower Hutt 5046, New Zealand

[5]Aalto University, Foundation sr, PO Box 11000, FI-00076 AALTO, Finland

*E-mail: evgeny.talantsev@imp.uran.ru

**Supplementary Table S1.** Deduced superconducting parameters for MATBG sample M1.

| Model | $\xi_{ab}(0)$ (nm) | $T_c$ (K) |
|---|---|---|
| Gorter-Casimir [S1,S2] | $63.7 \pm 0.7$ | $0.488 \pm 0.004$ |
| WHH [S3,S4] | $60.6 \pm 1.0$ | $0.503 \pm 0.007$ |
| B-WHH [S5] | $59.7 \pm 0.4$ | $0.497 \pm 0.003$ |
| JHC [S6] | $58.3 \pm 0.6$ | $0.515 \pm 0.005$ |
| Gor'kov [S7] | $61.9 \pm 0.4$ | $0.494 \pm 0.003$ |
| Eq. 18 [S8] | $64.0 \pm 0.8$ | $0.504 \pm 0.007$ |
| Average | $61.4 \pm 1.7$ | $0.500 \pm 0.006$ |

**Supplementary Table S2.** Deduced $I_c(sf,T)$ from fit of $V(I)$ curves to Eq. 23.

| Temperature (K) | $I_c(sf,T)$ (nA) | Uncertainty in deduced $I_c(sf,T)$ (nA) | *n-value* in Eq. 1 | Uncertainty in deduced *n-value* |
|---|---|---|---|---|
| 0.07 | 53.9 | <0.01 | 48.2 | <2 |
| 0.07 | 55.0 | <0.01 | 61.6 | <2 |
| 0.11 | 54.9 | <0.01 | 58.8 | <2 |
| 0.11 | 53.8 | <0.01 | 45.4 | <2 |
| 0.25 | 53.4 | <0.01 | 45.3 | <2 |
| 0.25 | 54.3 | <0.01 | 58 | <2 |
| 0.29 | 53.1 | <0.01 | 45 | <2 |
| 0.29 | 54.1 | <0.01 | 56.9 | <2 |
| 0.31 | 52.9 | <0.01 | 52 | <2 |
| 0.31 | 53.8 | <0.01 | 51.1 | <2 |
| 0.35 | 52.6 | <0.01 | 38 | <2 |
| 0.35 | 53.5 | <0.01 | 43.6 | <2 |
| 0.43 | 51.7 | <0.01 | 45.2 | <2 |
| 0.43 | 52.5 | <0.01 | 43 | <2 |
| 0.51 | 50.6 | 0.01 | 31.9 | 0.3 |
| 0.51 | 51.2 | 0.01 | 37.9 | 0.9 |
| 0.69 | 45.7 | 0.06 | 16 | 0.4 |
| 0.69 | 45.9 | 0.08 | 17.3 | 0.4 |
| 0.77 | 41.9 | 0.06 | 11.1 | 0.2 |
| 0.77 | 41 | 0.1 | 10.6 | 0.4 |
| 0.88 | 33.0 | 0.3 | 5.4 | 0.2 |
| 0.88 | 33.5 | 0.3 | 5.5 | 0.2 |
| 0.99 | 24.3 | 1 | 2.8 | 0.2 |
| 0.99 | 25.3 | 2 | 3.2 | 0.2 |
| 1.07 | 17.4 | 2 | 1.9 | 0.2 |
| 1.07 | 18.4 | 2 | 2 | 0.2 |
| 1.26 | 7 | 31 | 1.1 | 0.5 |
| 1.26 | 7 | 47 | 1.1 | 0.7 |

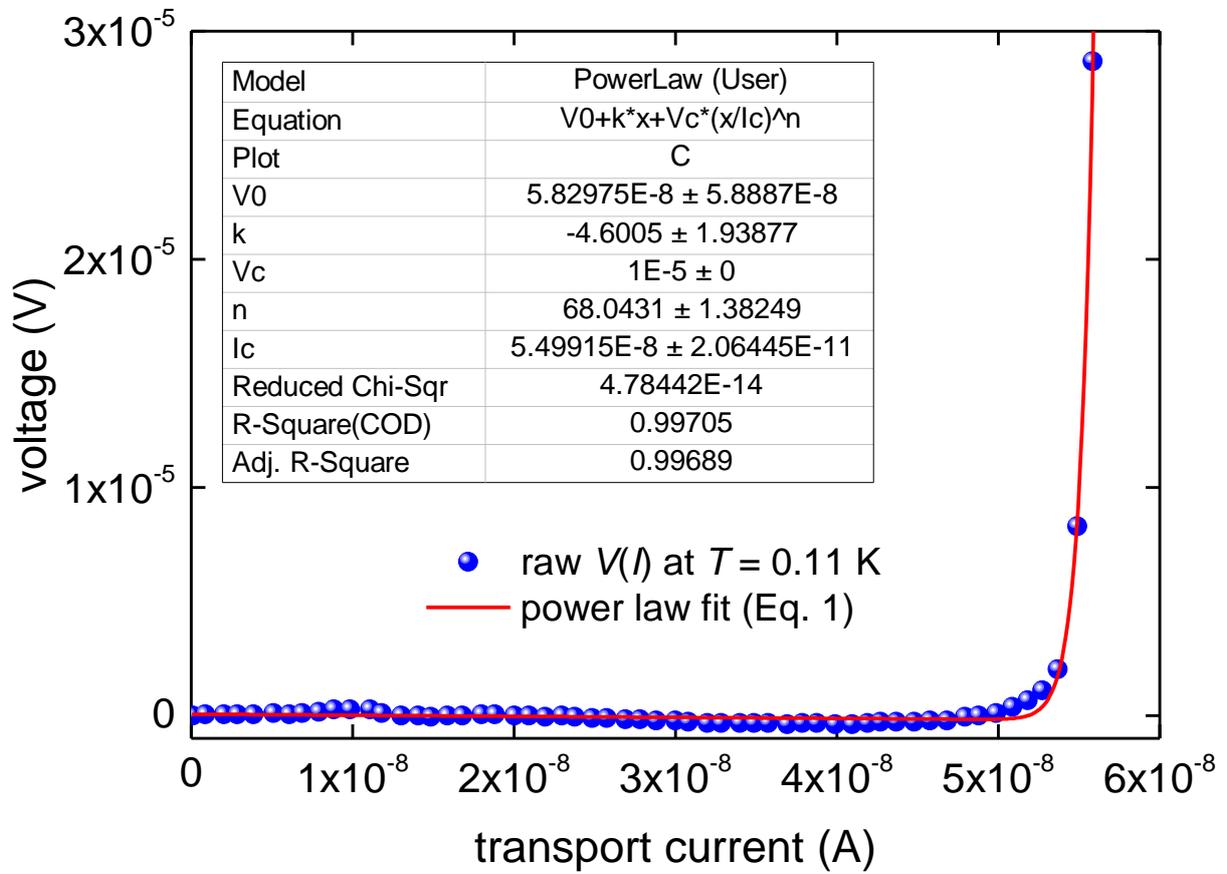

**Supplementary Figure S1.** Raw $V(I)$ curve at $T = 0.11$ K and data fit to Eq. 22.

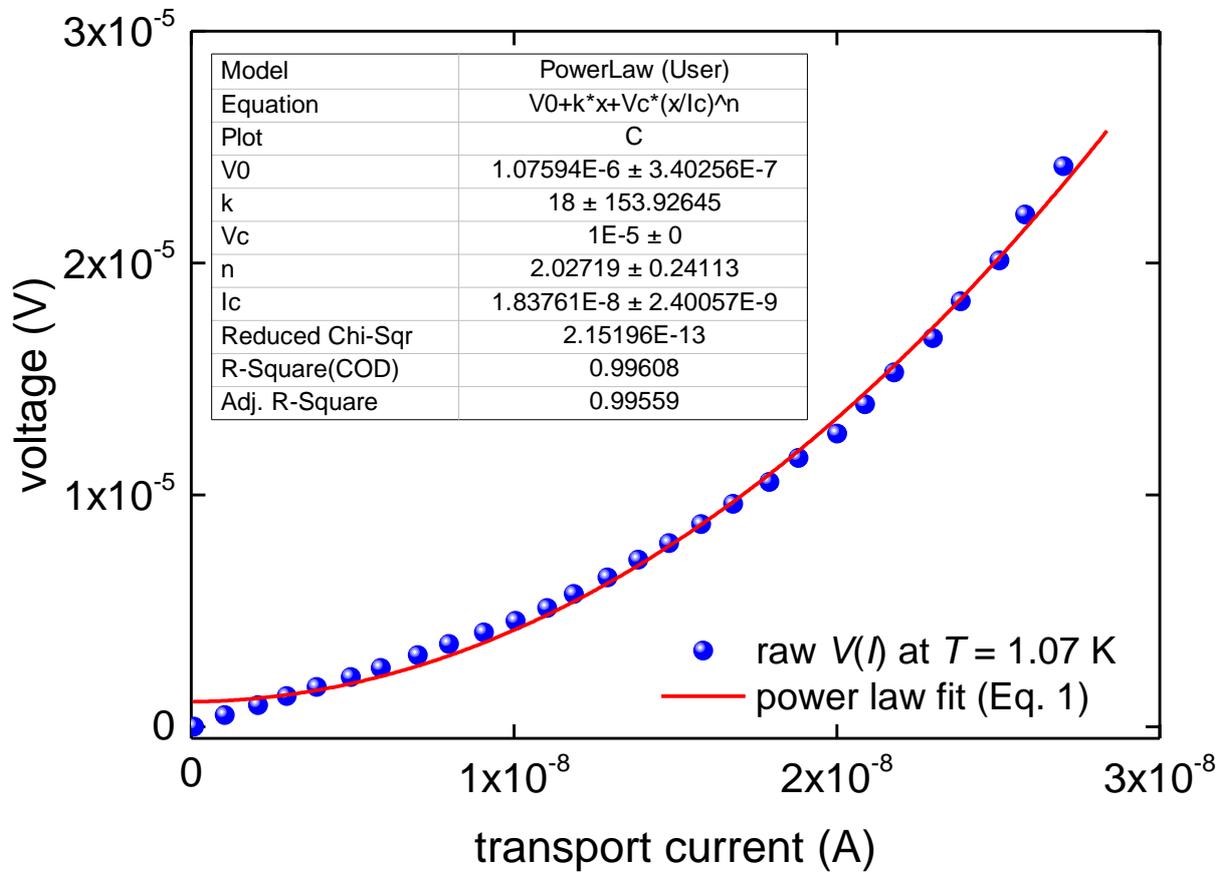

| Model | PowerLaw (User) |
| --- | --- |
| Equation | V0+k*x+Vc*(x/Ic)^n |
| Plot | C |
| V0 | 1.07594E-6 ± 3.40256E-7 |
| k | 18 ± 153.92645 |
| Vc | 1E-5 ± 0 |
| n | 2.02719 ± 0.24113 |
| Ic | 1.83761E-8 ± 2.40057E-9 |
| Reduced Chi-Sqr | 2.15196E-13 |
| R-Square(COD) | 0.99608 |
| Adj. R-Square | 0.99559 |

- raw $V(I)$ at $T = 1.07$ K
- power law fit (Eq. 1)

**Supplementary Figure S2.** Raw $V(I)$ curve at $T = 1.07$ K and data fit to Eq. 22.

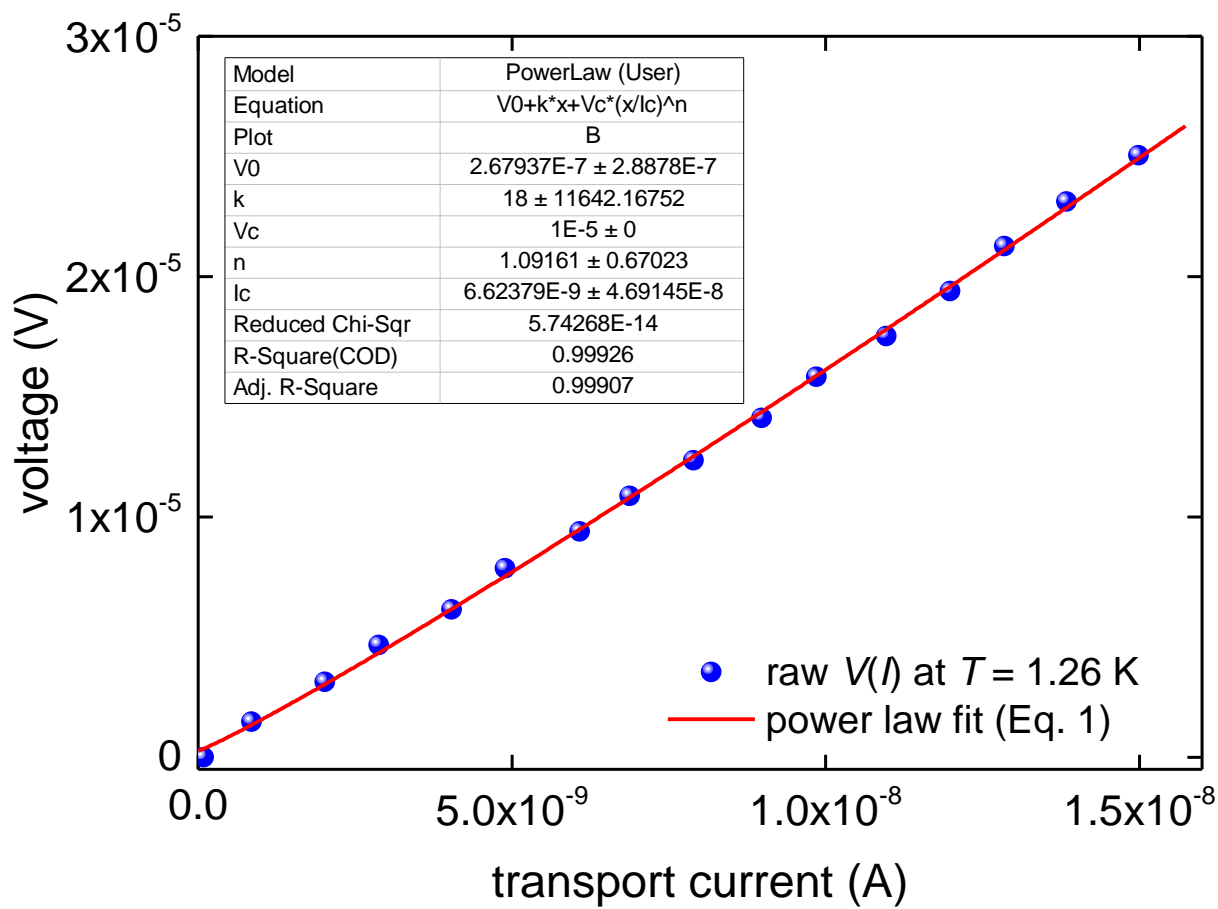

**Supplementary Figure S3.** Raw $V(I)$ curve at $T = 1.26$ K and data fit to Eq. 22.

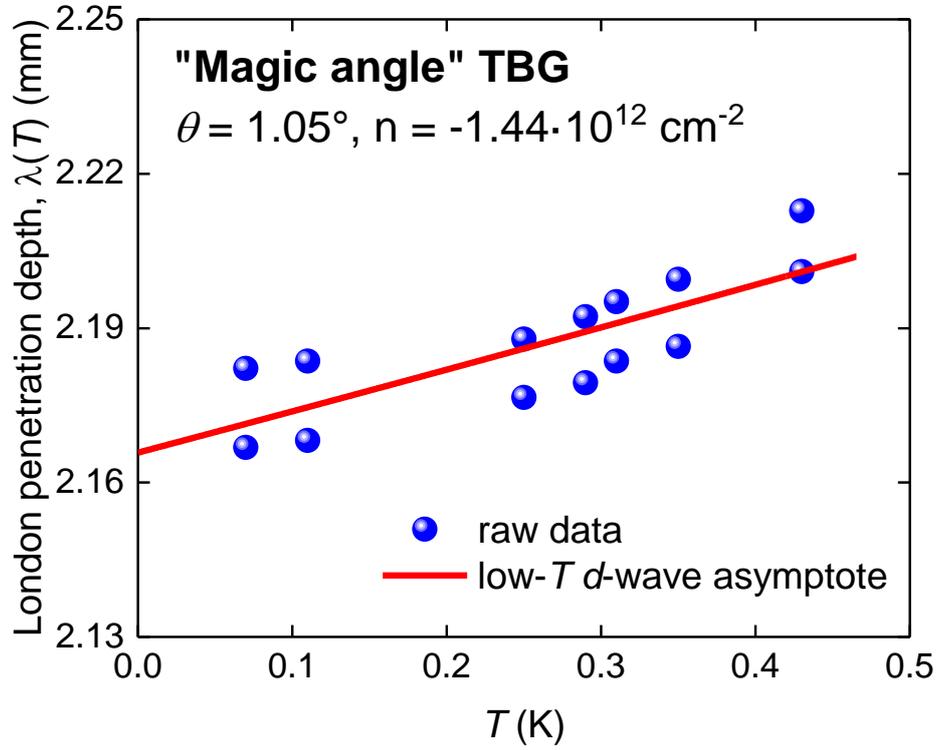

**Supplementary Figure S4.** London penetration depth, λ(T), for sample M2 (θ = 1.05°) of work of Cao et al [7] and data fit to low-temperature asymptote of *d*-wave model (see details in main paper and Table III). For this model we used κ = 35.6. The goodness of fit *R* = 0.6352.

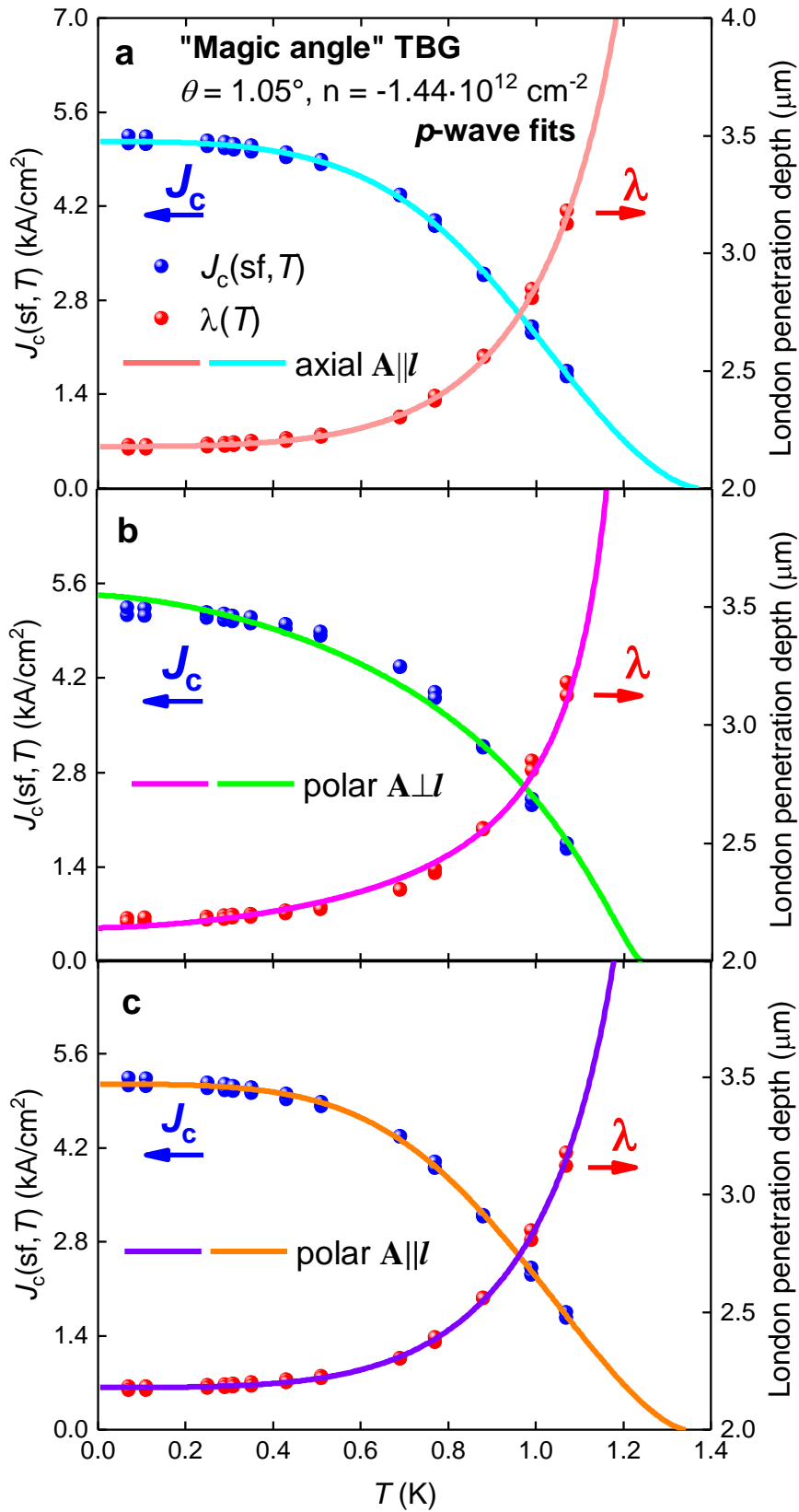

**Supplementary Figure S5.** The self-field critical current density, $J_c(sf,T)$, for sample M2 ($\theta$ = 1.05°) of work of Cao et al [7] and data fit to three cases of *p*-wave model (see details in main paper and Table III). For all models we used $\kappa$ = 35.6. (a) axial **A**||*l* fit, the goodness of fit $R$ = 0.9979; (b) *p*-wave polar **A**⊥*l* fit, $R$ = 0.9800; (c) polar **A**||*l* fit, $R$ = 0.9977.

## Supplementary References